# Effect of Substrate Surface on the Wide Bandgap SnO$_2$ Thin Films Grown by Spin Coating

Md. Nur Amin Bitu[1,3] · Nazmul Islam Tanvir[1,2,3] · Suravi Islam[1,3] · Syed Farid Uddin Farhad[1,2,3]

**Graphical Abstract**

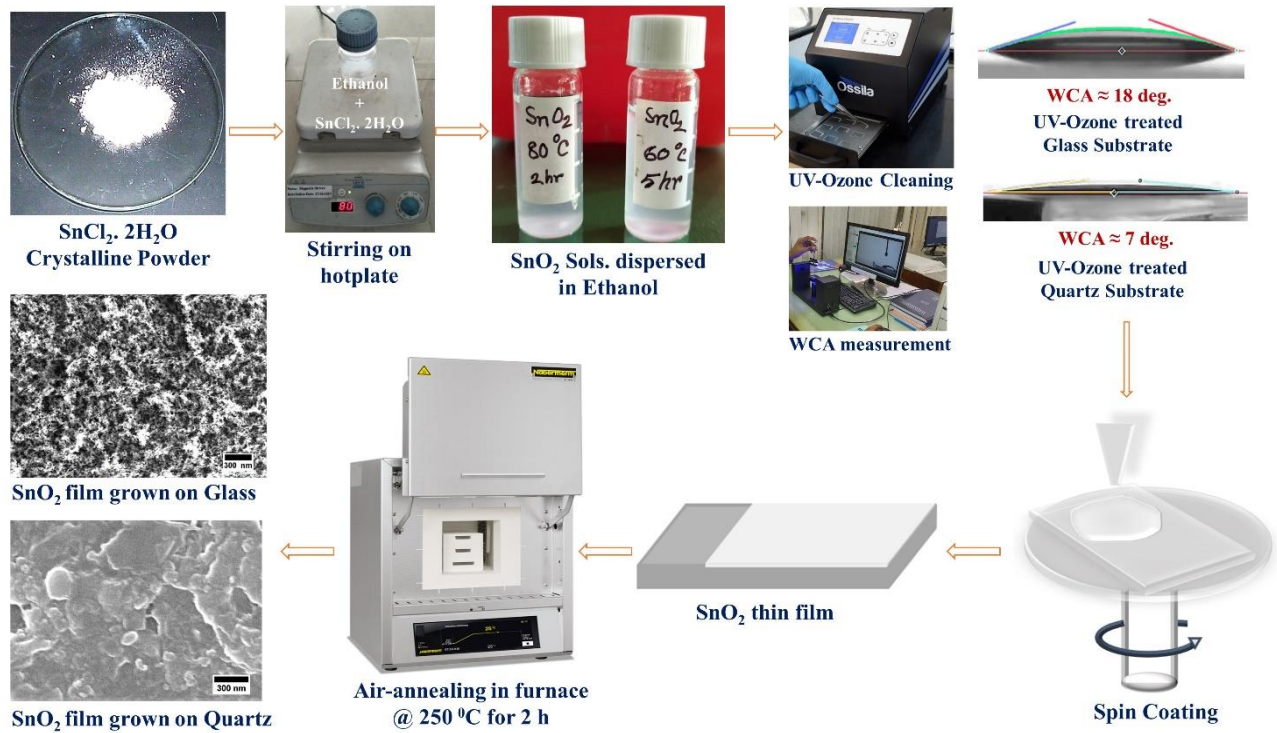

**Abstract**

Tin (IV) oxide (SnO$_2$) sols have been synthesized from SnCl$_2$.2H$_2$O precursor solution by applying two different processing conditions. The prepared sols were then deposited on UV-Ozone treated quartz and soda lime glass (SLG) substrates by spin coating. The as-synthesized film was soft-baked at about 100 °C for 10 min. This process was repeated five times to get a compact film, followed by air-annealing at 250 °C for 2 hours. The pristine and annealed films were characterized by UV–Vis-NIR spectroscopy, Grazing Incident X-Ray Diffraction (GIXRD), and Field Emission Scanning Electron Microscope (FESEM). The effect of the substrate surface was investigated by measuring the contact angles with De-Ionized (DI) water. UV-Ozone treatment of substrate provides a cleaner surface to grow a homogeneous film. The electrical resistivity of annealed thin films was carried out by a four-point-collinear probe employing the current reversal technique and found in the range of ~2×10$^3$ to 3×10$^3$ Ω.cm. Film thickness was found in the range of ~137 - 285 nm, measured by a stylus profilometer. UV-VIs-NIR Transmission data revealed that all the thin film samples showed maximum (82 - 89) % transmission in the visible range. The optical bandgap of the thin films was estimated to be ~3.75 - 4.00 eV and ~3.78 - 4.35 eV for the films grown on SLG and quartz substrates, respectively.

✉ Syed Farid Uddin Farhad
sf1878@my.bristol.ac.uk;  s.f.u.farhad@bcsir.gov.bd

[1] Energy Conversion and Storage Research Section, Industrial Physics Division, BCSIR Dhaka Laboratories, Dhaka 1205, Bangladesh

[2] Central Analytical and Research Facilities (CARF), BCSIR, Dhaka 1205, Bangladesh

[3] Bangladesh Council of Scientific and Industrial Research (BCSIR), Dhaka 1205, Bangladesh

## Introduction

Metal oxide semiconductors are intriguing because of their tunable optical bandgap, variable physicochemical properties, nature of conductivity, excellent mechanical and chemical stability [1–3], etc. With the help of advanced technology, as it is possible to produce metal oxides as thin films [4], nanoparticles [5], nanowires, and nanorods [6], their application in semiconductor electronics and energy conversion and storage devices has increased over time. Thin films of such metal oxides are desirable as electrodes for optoelectronic devices that need materials that can be transparent to light and conduct electricity. These thin films enable light to flow through with minimum absorption, leading to transparent conductive oxide (TCO) [7]. Due to their conducive electro-optical properties, highly transparent and conductive oxide thin films are extensively used materials, especially in solar cells. TCOs such as tin (IV) dioxide ($SnO_2$) thin films with wide bandgap energies ($E_g$ >3.4 eV) have tremendous interest in transparent conducting electrodes for displays, n-type transport layers for high-efficiency perovskite solar cells, as well as other optoelectronic devices [8–10]. $SnO_2$ thin films can be synthesized using different deposition techniques. M. Rahayi et al. [11] have synthesized $SnO_2$ thin films on glass substrates via an RF magnetron sputtering and found the optical bandgap 3.18 – 3.21 eV for different annealing temperatures (300 °C, 400 °C and 500 °C). This bandgap is much lower than our desired wide bandgap $SnO_2$ thin film ($E_g$ >3.4 eV). F.R. Chowdhury et al. [12]. reported an optical bandgap of thermally evaporated $SnO_2$ thin films in the range of 3.38 – 3.59 eV. V. Janakiraman et al. [13] reported that $SnCl_2 \cdot 2H_2O$ precursor solution derived $SnO_2$ thin films deposited onto microscopic glass substrates at 350 °C by spray pyrolysis [13]. From the UV visible absorption spectra, they have found that the optical bandgap of $SnO_2$ thin film is 3.57 eV. L. lin et al. [14] reported the $SnCl_2 \cdot 2H_2O$ precursor solution derived wide bandgap (3.91 – 4.03 eV) $SnO_2$ thin films grown onto FTO/glass substrate by spin coating technique. Another wide bandgap (4.1 eV) $SnO_2$ thin films grown on a double-polished r-plane sapphire substrate using plasma-assisted molecular beam epitaxy reported by T. Oshima et al. [15]. In all these works, $SnO_2$ thin films were grown on glass or other substrates at relatively higher temperatures (≥300 °C). However, the high processing temperature is sometimes not suitable for flexible or soda-lime glass substrates (glass softening temperature is ~300 °C ([6] and refs. therein)). Additionally, in the case of superstrate solar cells or other optoelectronic devices, high growth temperatures may degrade the quality of the underlying layer(s).

The surface nature of substrates plays a critical role in nucleation, coalescence of islands, and subsequent film growth that dictate their overall surface roughness and morphology, resulting in their optoelectronic properties. To this end, we choose the two most popular non-conducting amorphous substrates, namely SLG and Quartz, to investigate their effect on the growth of $SnO_2$ films and their structural, optical, electrical, and morphological properties. The annealing temperature of 250 °C was chosen intentionally to prevent the possible micro-cracking of both deposited layers on the substrate surface as well as the SLG substrate itself. In addition, the use of amorphous quartz substrates facilitated us in finding the actual bandgap of our synthesized $SnO_2$ thin films, which are discussed below.

## Materials and methods

### Substrate Preparation

Before thin film deposition, soda lime glass (SLG) and quartz substrate of (12×12×1) $mm^3$ size were initially scrubbed with detergent soap in clean tap water and then immersed into a hot water-filled beaker for ultrasonic cleaning. To remove the grease and other stubborn dirt from the surface of the substrate, they were further subjected to ultrasonic cleaning sequentially in toluene, acetone, isopropanol, and finally in deionized (DI) water at 40 °C, each step for 15 min. Then, to eliminate moisture from the substrate's surface, they were dried by a compressed air-blower and air-annealed at 100 °C for 5 min. Finally, the substrates were subjected to UV-ozone treatment for 20 min to remove organic residues from their surface just before the spin coating.

### Thin Film Fabrication

Tin (II) chloride di-hydrate ($SnCl_2.2H_2O$, purity ~98 %, Alfa Aesar) was used as a starting material for preparing the precursor solution. Briefly, tin (II) chloride di-hydrate (molar mass: 225.65 g/mol) of 0.5641 g was dissolved with 25 ml ethanol ($C_2H_5OH$, purity ~99.9 %, Merck Germany) to prepare 0.1 M precursor solutions. Then the solution was stirred at 500 rpm with a magnetic stirrer coupled hotplate under two conditions (80 °C for 2 h and 60 °C for 5 h) to synthesize $SnO_2$ Sols. The solution was then filtered using a 0.20 μm polytetrafluoroethylene (PTFE) filter before coating the substrates. Spin-coating was carried out using a commercial spin coater (Polos-SPIN150i-NPP, Germany) with two steps (500 rpm, 5 s, and 2000 rpm, 25 s). The prepared film was soft-baked at about 100 °C for 10 min. This process was repeated five times to get five layers for each sample to achieve the desired film thicknesses. After deposition, $SnO_2$ thin films were air-annealed at 250 °C for 2 h in a furnace (model: Nabertherm P-310, Germany) with a heating rate of 5 °C/min. The $SnO_2$ thin films grown using the solution synthesized at 80 °C for 2 h and 60 °C for 5 h on glass substrates were abbreviated as G-1 & G-2, and those on quartz are abbreviated as Q-1 & Q-2, respectively.

## Thin Film Characterizations

The optical transmission and absorption data were recorded at room temperature by a UV–Vis–NIR spectrophotometer (SHIMADZU 2600, Japan) in the wavelength range of 220 – 1400 nm. The structural characterization of the deposited thin films was performed by the grazing incident x-ray diffraction (GIXRD) technique using a Thermo Scientific ARL EQUINOX 1000 X-Ray Diffractometer (USA) with a monochromatic copper anode source of Cu-K$\alpha$ radiation, $\lambda$ = 1.5406 Å in coplanar asymmetric geometry. The thickness of all films was measured by an Alpha-Step D-500 stylus profilometer. The electrical resistivity of the films was measured using a four-point collinear probe (Ossila, UK) coupled with a Keithley 2450 Source Measure Unit (SMU). The surface morphology of the deposited films was investigated using a Field Emission Scanning Electron Microscope (FESEM) (JEOL, JSM-IT800). A contact angle goniometer, coupled with a high-resolution video camera (Ossila, UK), was used to measure the blank substrates' water contact angle (WCA) before and after cleaning with different solvents and UV-Ozone treatment.

## Results and Discussion

### Effect of Substrate Cleaning by Water Contact Angle Measurement

A properly cleaned substrate surface is the prerequisite for homogenous and compact thin film growth. To assess the degree of cleanliness of substrate surfaces by the procedure described in the substrate preparation section, all the cleaning steps were systematically investigated by the water contact angle (WCA) measurements (Fig. 1). The experiment shows that the WCA of quartz without cleaning decreases a little from 24.15 ± 1.55 deg. to 16.35 ± 1.43 deg. within 6 s (Fig. 1b). In contrast, after sonication with different solvents, WCA decreases from 9.60 ± 0.37 deg. to 5.06 ± 0.11 deg. within 200 ms. When the substrate was further subjected to UV-ozone treatment, the 185 nm radiation generated from the UV-Ozone cleaner reacted with air-oxygen ($O_2$) to produce ozone ($O_3$), which decomposed and transformed organic pollutants on the surface of the substrate into volatile substances, which can be removed easily [16]. Therefore, after UV-Ozone cleaning WCA of quartz decreases significantly from 7.14 ± 0.12 deg. to 1.82 deg. within 100 ms (Fig. 1c). In the case of SLG substrate, the WCA remains almost constant or decreases very slowly (Fig. 1d). From this observation, we can conclude that UV-Ozone treatment offers better cleaning results on quartz than on SLG substrate, which may be beneficial for producing a homogeneous thin film.

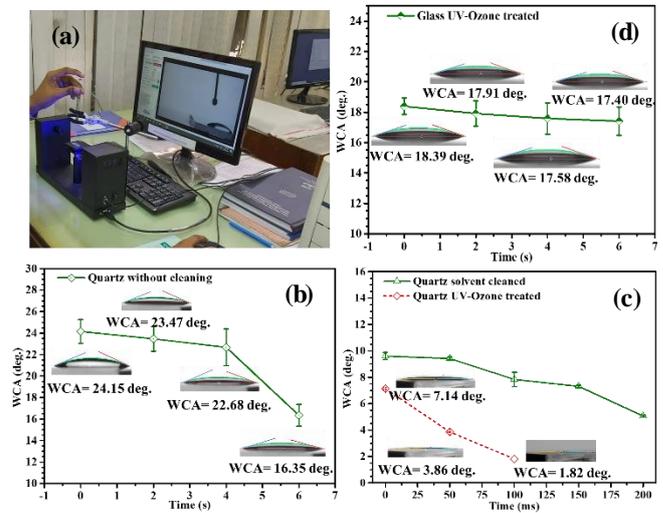

**Fig. 1** Water contact angle (WCA) measurements to show the hydrophilic nature of cleaned Glass and Quartz substrates' surface.

### Optical Properties

Results of all the optical properties of $SnO_2$ thin films are depicted in Fig. 2. The absorption spectra of the synthesized $SnO_2$ sols are also shown in Fig. 2a. A strong absorption peak that appeared near about 244 nm and 249 nm confirms the formation of $SnO_2$ sols [17].

The optical transmission spectra of all the thin films, along with the blank quartz and SLG substrate, are shown in Fig. 2b. The transmission data revealed that all the films are highly transparent (82-89 % transmission near 535 nm) in the visible region. It is worth noting that the same sols were deposited on quartz and SLG substrates. Here, we observed that the SLG substrate absorbs light near about 300-350 nm range, but quartz doesn't, which facilitates finding the actual transmission spectra and the absorption edge of $SnO_2$ thin films.

The direct bandgap of the $SnO_2$ thin films was found by extrapolating the linear portion of the graph of $(\alpha h\nu)^2$ versus photon energy ($h\nu$) as shown in Fig. 2c and Fig. 2d for before and after annealing, respectively. The results showed that the bandgap of $SnO_2$ films was found to be around (3.75 to 4.00) eV and (3.78 to 4.35) eV for thin film grown on SLG and quartz substrates, respectively. Since quartz does not absorb UV light up to 350 nm like SLG (see arrow mark in Fig.2b), therefore, quartz substrates allowed us to observe the actual wide bandgap (3.78 to 4.35) eV of our synthesized $SnO_2$ thin films [14, 15].

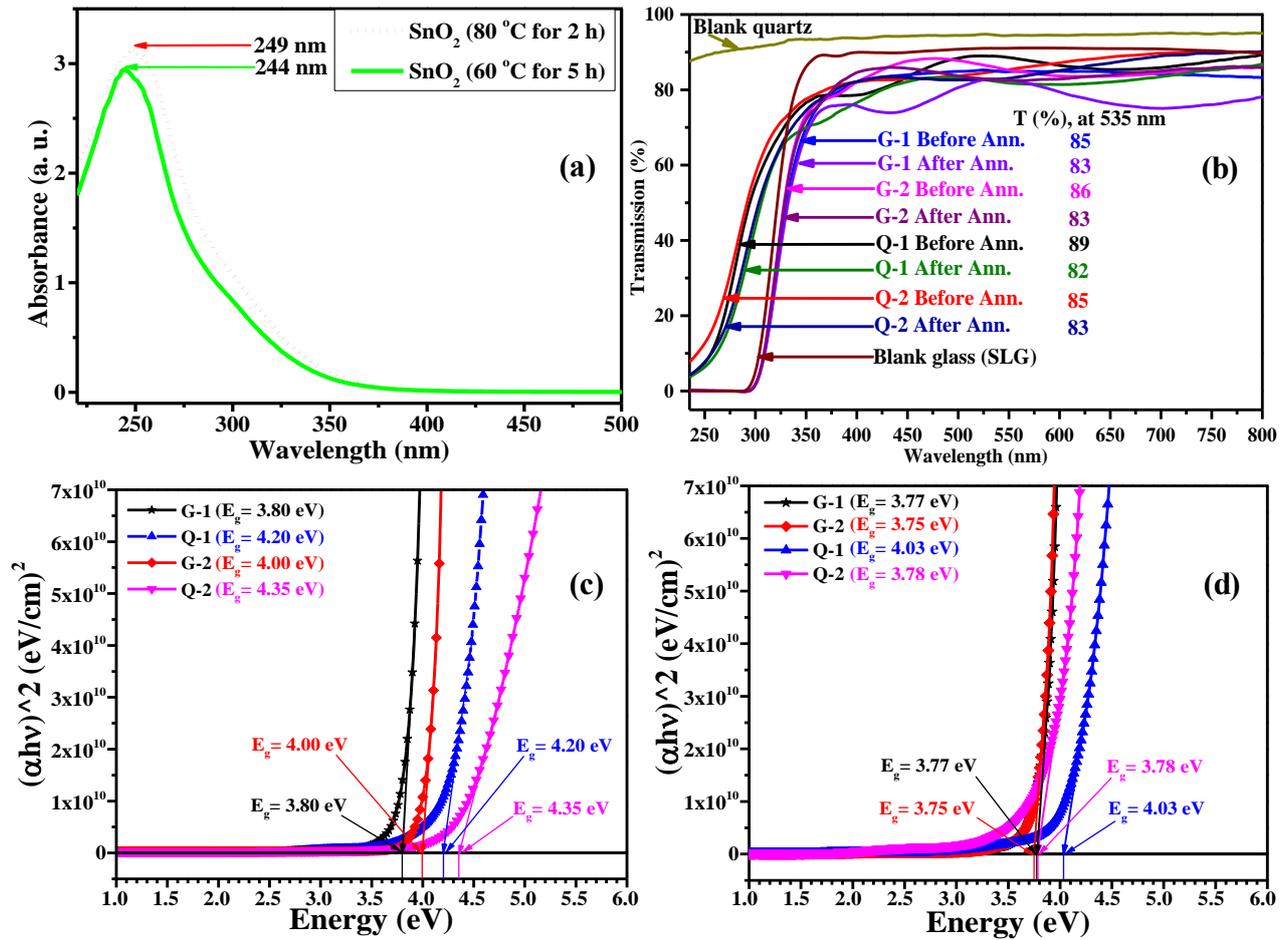

**Fig. 2 a** Absorption spectra of the synthesized $SnO_2$ sols. **b** Optical transmission spectra of $SnO_2$ thin films before and after annealing. **c** Estimated direct bandgap of $SnO_2$ thin films before annealing and **d** After annealing.

### Electrical Properties and Thickness measurement

The thickness of the $SnO_2$ films was measured using a stylus profilometer. Results showed that the film thicknesses are 285 nm, 238 nm, 276 nm and 137 nm for G-1, G-2, Q-1, and Q-2 films, respectively. The electrical resistivity of the films grown on insulating (quartz and glass) substrates was measured carefully with the current-reversal technique with the best forcing currents to avoid the thermal contribution in the measurements [3]. The electrical resistivity of the thin films G-1, G-2, Q-1 and Q-2 measured as $2.96\times10^3$ Ω-cm, $2.25\times10^3$ Ω-cm, $2.09\times10^3$ Ω-cm and $2.01\times10^3$ Ω-cm respectively. Note that the electrical resistivities of thin films produced on quartz are lower than those grown on SLG substrates. This presumably is due to more compact film from easy nucleation on the cleaner quartz substrate surface.

### Structural Properties

The crystalline phase and structure of the synthesized $SnO_2$ thin films were confirmed by GIXRD (Fig. 3). For $SnO_2$ thin film, G-1 (Fig. 3a), the angle of incidence was set from 3.0 to 10.0 deg. with an increment of 0.5 deg. For $SnO_2$ thin film, Q-1 (Fig. 3b), the angle of incidence was set as a program of two steps; one is 0.1 to 0.5 deg., and another is 0.5 to 4.5 deg. with an increment of 0.1 deg. and 0.5 deg., respectively. GIXRD patterns of Q-1 exhibited a better XRD pattern from 0.3 deg. to 2.0 deg. and that of G-1 showed a better XRD pattern from 3.0 deg. to 5.0 deg. of incident angle (cf. Fig. 3a and Fig. 3b). Notice a slight shift of Bragg peak position of $SnO_2$ in GIXRD patterns with the increasing grazing incidence angles is due to the layer's penetration depth variation experienced by the probing X-ray beam. This is more conspicuous in Fig. 3b. where the intensity of quartz-hump (dashed line) and the $SnO_2$ (110) peak (solid line, taken only first peak for simplicity) is, respectively, seen to be increasing and decreasing with the increasing incident angle from 0.1 deg. to 4.5 deg. consistently. The apparent peak shifting with the variation of grazing incident angles observed in GIXRD patterns is due to the incident X-ray beam's refraction and the sample position's misalignment in the coplanar asymmetric geometry [18], which is significantly different from the symmetric Bragg-Brentano XRD

(BBXRD) geometry. Therefore, prominent broadening of Bragg peaks may arise in GIXRD patterns even for the well-crystalline samples [19] due to the width variation of the X-ray beam footprint on the sample under study [18]. Nevertheless, GIXRD is a very useful tool to investigate different crystalline information of ultra-thin specimens, which is often impossible in BBXRD. As can be seen from the GIXRD patterns in Fig. 3, observed broad peaks diffracting from (110), (101), (211), and (112) planes indicate the characteristic peaks of tetragonal $SnO_2$ [20] for all samples. A comparative study on the best GIXRD patterns for the film on the quartz substrate (Q-1) and the glass substrate (G-1), recorded respectively with grazing incidence angles 3 and 0.5 deg., is shown in Fig. 3c. The full width at half maximum (FWHM) value of (211) diffraction peak is ~ 3.07 deg. for G-1 and ~ 3.69 deg. for Q-1 samples, which implies an average crystallite domain size of ~ 2.86 nm for G-1 and ~2.39 nm for Q1 using Scherrer formula [21]. Therefore, one must take special care in estimating crystallite size (i.e., the size of coherent domains) from GIXRD patterns [22]; otherwise, it may result in an underestimation of crystallite size.

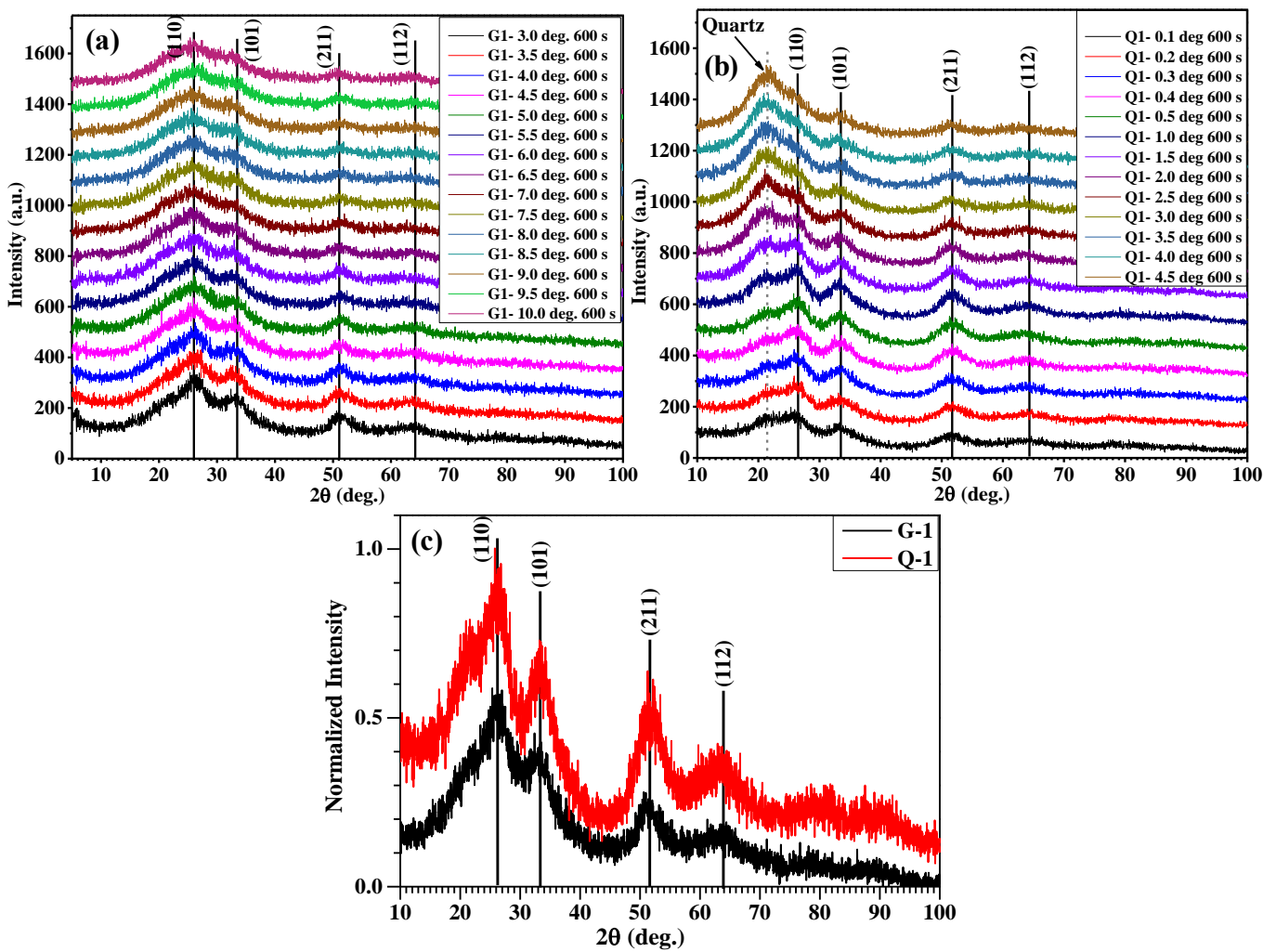

**Fig. 3** GIXRD patterns of $SnO_2$ thin films grown on SLG (G-1) and Quartz (Q-1) substrates using similar processing conditions.

## Morphological Properties

The surface morphologies of typical $SnO_2$ thin films on SLG (G-2) and quartz (Q-2) are shown in Fig. 4. The FESEM micrographs show that $SnO_2$ thin films deposited on both substrates are compact, but the film on the quartz substrate appears with some agglomeration. The average particle size was estimated by ImageJ software, and it was found to be 24 ± 3 nm for G-2 and 27 ± 4 nm for Q-2. (see bottom panel of Fig. 4.)

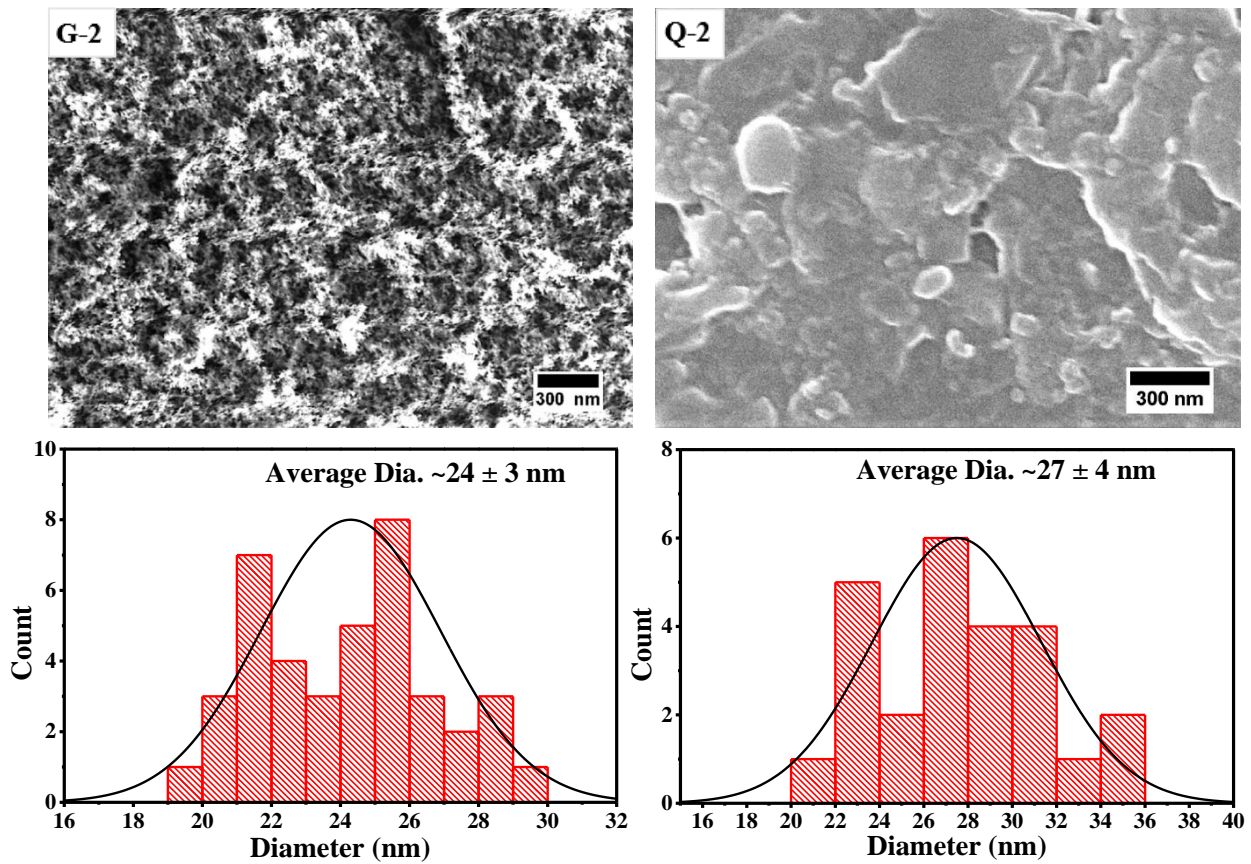

**Fig. 4** FESEM micrographs (top) and grain size distribution curves (bottom) of $SnO_2$ thin films grown on SLG (G-2) and Quartz (Q-2) substrates using similar processing conditions.

## Conclusions

In summary, highly transparent and wide bandgap thin $SnO_2$ films were successfully grown on UV-Ozone treated quartz and SLG substrate from $SnCl_2 \cdot 2H_2O$ precursor solution by spin coating. UV-VIs-NIR spectra revealed that all the samples with thickness ~137 - 285 nm showed maximum 82 - 89 % transmission in the visible region. The estimated bandgap from transmission spectra was found in the range of 3.75 - 4.00 eV and 3.78 - 4.35 eV for the thin films grown on SLG and quartz substrates, respectively. The use of quartz substrates facilitated confirmation of the exceptionally wide bandgap of $SnO_2$ thin films produced at a processing temperature as low as 250 $^0$C. The electrical conductivity of $SnO_2$ thin films grown on the quartz substrate was higher than those grown on the SLG substrate. GIXRD and FESEM showed the tetragonal phase and compact morphological features of the low-temperature processed $SnO_2$ thin films. In this study, we also demonstrated that the UV-Ozone treatment of substrate provides a cleaner surface than any other conventional cleaning process to grow a homogeneous film assessed by lowering the water contact angles.

**Acknowledgments** All the authors gratefully acknowledge the experimental support of the Energy Conversion and Storage Research (ECSR) Section, Industrial Physics Division, BCSIR Laboratories, Dhaka 1205, Bangladesh Council of Scientific and Industrial Research (BCSIR), under the scope of R&D project # 03-FY2017-2022. S.F.U. Farhad acknowledges the support of TWAS grant # 20-143 RG/PHYS/AS_I for ECSR, IPD.

**Conflict of interest** On behalf of all authors, the corresponding author states that there is no conflict of interest.

**Data Availability** Data will be made available on reasonable request.